\documentclass[12pt,preprint]{aastex}

\usepackage{float}
\usepackage{multicol}

\begin{document}

\title{Mantle Convection, Plate Tectonics, and Volcanism on Hot Exo-Earths}
\author{Joost van Summeren}
\email{summeren@hawaii.edu}
\author{Clinton P. Conrad}
\author{Eric Gaidos}
\affil{Department of Geology and Geophysics, University of Hawai'i at M${\bar \textrm{a}}$noa, Honolulu, Hawai'i 96822, U.S.A.}

\begin{abstract}
Recently discovered exoplanets on close-in orbits should have surface temperatures of 100's to 1000's of K.
They are likely tidally locked and synchronously rotating around their parent stars and, if an atmosphere is absent, have surface temperature contrasts of many 100's to 1000's K between permanent day and night sides.
We investigated the effect of elevated surface temperature and strong surface temperature contrasts for Earth-mass planets on the
(i) pattern of mantle convection,
(ii) tectonic regime, and
(iii) rate and distribution of partial melting, using numerical simulations of mantle convection with a composite viscous/pseudo-plastic rheology.
Our simulations indicate that, if a close-in rocky exoplanet lacks an atmosphere to redistribute heat, a $\gtrsim$ 400 K surface temperature contrast can maintain an asymmetric degree 1 pattern of mantle convection in which the surface of the planet moves preferentially toward subduction zones on the cold night side.
The planetary surface features a hemispheric dichotomy, with plate-like tectonics on the night side and a continuously evolving mobile lid day side with diffuse surface deformation and vigorous volcanism.
If volcanic outgassing establishes an atmosphere and redistributes heat, plate tectonics is globally replaced by diffuse surface deformation and volcanism accelerates and becomes distributed more uniformly across the planetary surface.
\end{abstract}


\begin{multicols}{2}[.]
\section{Introduction}
Recent discoveries of exoplanets with Earth-like mass and radius \citep{Mayor_etal2009,Howard_etal2010,Borucki_etal2011} have intensified debate on how such planets compare to Earth in other respects.
Due to observational bias, many of the discovered exoplanets inhabit short-period, close-in orbits and have effective temperatures exceeding many hundreds of Kelvin.
For such planets, a likely outcome of dynamical evolution is tidal locking to their parent stars and, most probably, capture into a 1:1 spin-orbit resonance \citep{CorreiaLaskar2010b}.
Synchronous rotation causes asymmetric insolation and, in the absence of a substantial atmosphere, a strong (100's to 1000's K) temperature contrast between these planets' permanent day and night sides, as has been estimated for CoRoT-7b \citep{Leger_etal2009} and is plausible for Kepler-10b \citep{Batalha_etal2011}.

Characterization of exoplanet surfaces is challenging and numerical simulations can help determine possible scenarios.
Subsolidus convection is likely within the silicate mantles of rocky exoplanets but its vigor and surface expression depend on mantle temperature, composition, and rheology.
Within the Solar System, Earth is the only planet that currently exhibits plate tectonics which, on a geologic time scale, regulates volatile species in Earth's atmosphere via volcanic outgassing, silicate weathering, and subduction of precipitated carbon \citep{Walker_etal1981}.
Venus is thought to experience infrequent global-scale resurfacing events, possibly the result of mantle-wide episodic overturn and associated with the formation of a dense greenhouse atmosphere, e.g., \citet{Solomon_etal1999}.
Mars has experienced recent substantial volcanism although there is no evidence for recent crustal mobility, suggesting a stagnant lid regime with convective activity in the mantle interior below a static single-plate lithosphere, e.g. \citet{Spohn_etal2001}.
Elevated surface temperatures can affect a planet's interior dynamics \citep{Lenardic_etal2008} and its surface properties, as exemplified by insolation-driven variations of Mercury's lithospheric strength \citep{Williams_etal2011}.
Although there are no examples in our Solar System, a hemispheric contrast in surface temperature may influence the interior dynamics and surface expression of exoplanets.

Here, we investigate the effect of elevated surface temperature and strong surface temperature contrasts for Earth-mass planets on
(i) patterns of mantle convection,
(ii) tectonic regimes, and
(iii) the rate and distribution of partial melting (volcanism).
For this purpose, we conducted numerical simulations of mantle convection with imposed surface temperatures.

\section{Description of Numerical Model and Parameters}
\label{sct-method}
We investigated planetary mantle convection through numerical simulations of an incompressible fluid at infinite Prandtl number using the classical Boussinesq formulation.
The associated conservation equations for mass, momentum, and energy are solved numerically using the finite element package CitComS-3.1.1 \citep{Zhong_etal2000}.
We investigated effective Rayleigh numbers $Ra_{eff}=\rho_0 g_0 \alpha_0 \Delta T h^3 / (\eta_{eff} \kappa_0)$ in the range $\sim$10$^5-10^7$, where $\rho_0$, $g_0$, $\alpha_0$, and $\kappa_0$ are the respective reference values for density, gravitational acceleration, thermal expansivity, and thermal diffusivity, $\eta_{eff}$ is the time-averaged mantle viscosity, $\Delta T$ is the temperature contrast across the mantle, and $h$ is the mantle thickness.
Bottom and internal heating both contribute to the heat budget.
Internal radiogenic heat production is uniformly distributed across the mantle domain and constant in time in our models.
The non-dimensional internal heat generation rate, defined as $\gamma=h^2 H_0 / (\kappa c_P \Delta T)$ \citep{Glatzmaier1988}, is set to 11; this corresponds to a dimensional heating rate of $H_0 = 4 \times 10^{-12}$ W kg$^{-1}$, close to the present-day chondritic heating rate \citep{TurcotteSchubert2002}.

Our calculations ignore dissipation of tidal forces that can contribute to mantle heating and thermal runaway, depending on orbital eccentricity and type of resonance \citep{Behounkova_etal2011}.
However, without external perturbations, tidal dissipation renders orbital eccentricity insignificant on Myr--Gyr time scales and tidal effects can be neglected.

For the mantle domain, we employ an annulus (bi-section of a 3-D spherical domain)  that is aligned with the ecliptic plane.
All domain boundaries are free-slip and two side boundaries are imposed at the antistellar point.
We adopt Earth's outer radius ($r_S$ = 6371 km) and mantle thickness ($h$ = 2891 km), and allocate 257 x 65 nodal points in the lateral and radial directions, respectively, with gradual mesh refinement towards the top and bottom boundaries where temperature variations are generally greatest.

To permit plate-like behavior of the surface boundary layer, we adopt a composite viscous/pseudo-plastic rheology in our models, following \citet{Tackley2000b,Tackley2000c}.
Temperature-dependent viscosity $\eta_v$ is described by an Arrhenius-type law:

\begin{equation}
\eta_v(T') = \eta_0 \exp \left[23.03 \left(\frac{1}{1+T'} - \frac{1}{2} \right) \right],
\end{equation}

where the reference viscosity is $\eta_0 = 5 \times 10^{20}$ Pa s, and $T'$ is the dimensionless mantle potential temperature that relates to the dimensional temperature, $T$, as $T'=(T-T_S) /\Delta T$, with $\Delta T$ = 2400 K \marginpar{(12)}the potential temperature contrast across the mantle and $T_S$ the surface potential temperature.
Viscosity changes by 5 orders of magnitude over the considered temperature range ($T' \in [0,1]$) and generates a lithosphere over a weaker mantle.
Pseudo-plastic yielding concentrates strain and allows for lithospheric break-up in confined regions that mimic subduction zones and spreading centers \citep{Tackley2000b,Richards_etal2001}.
In regions where the model stress exceeds an assigned yield stress, $\sigma_y$, a yield viscosity is calculated as $\eta_y=\sigma_y/2\dot{\epsilon_{II}}$, where $\dot{\epsilon_{II}}$ is the second invariant of the strain rate tensor.
The composite rheology is described as $\eta = \min\left(\eta_v(T),\eta_y\right)$.
We did not consider strain- or strain-rate-weakening \citep{Christensen1984}, or time-dependent damage rheology \citep{Bercovici1996}, which could further enhance plate-like behavior but would significantly complicate our analysis.

We assign distinct temperature conditions for 3 contrasting cases.
In a first set ("cold" or C-models) we apply a uniform, time-constant surface temperature $T_S$ = 273 K, comparable to Earth.
A second set ("hot" or H-models) mimics close-in planets (i.e., orbital distance of $a$ = 0.13 AU around a solar mass star) with efficient heat redistribution and a uniform surface temperature of $T_S$ = 759 K.
A third set ("asymmetric" or A-models) considers planets on a similar close-in orbit that lack heat redistribution.
For these models, a high substellar temperature of $T_{subst}$ = 1073 K decreases sinusoidally to the terminus and is kept constant on the night side at $T_S$ = 273 K.
For all models, the core-mantle boundary (CMB) potential temperature is uniform and constant at $T_{CMB}$ = 2673 K.
Each simulation is run for several billion years (Gyr) of model time and we exclude the first 1 Gyr of initial transients to focus on statistically steady-state behavior.

To quantify tectonic regimes, we make use of two previously-defined diagnostics \citep{Tackley2000b}.
First, to quantify the localization of surface strain rates, we define "plateness" as $P = 1 - (f_{80}/0.6)$, where $f_{80}$ is the area fraction that encompasses 80\% of the total surface strain rate.
$P=0$ corresponds to strain localization for isoviscous convection.
Second, lid mobility $M$ is defined as the ratio of the root-mean-square (RMS) surface flow velocity relative to the RMS velocity of the entire mantle domain, $M=v_{RMS}^{srfc} / v_{RMS}^{whole}$.
Models with plate-like behavior are characterized by $M \approx 1-1.5$ and for stagnant lid convection $M\sim0$.
To allow faster calculations, we made use of the symmetry of the problem and determined tectonic regimes (Section \ref{sct-res1}) for models of 180$^o$ opening angle, with side boundaries at the sub-stellar and anti-stellar points.
For 4 representative models the time-averaged $P$ and $M$ values differ by only $6$\% and $9$\%, respectively, when comparing models of 180$^o$ and 360$^o$.

Pressure-release partial melting is calculated following \citet{Raddick_etal2002}.
We only consider melting in regions where convective flow is upward and where real (potential + adiabatic) temperatures exceed the mantle solidus temperature ($T_r>T_{sol}$).
For $T_{sol}$ we use a parameterization for dry mantle peridotite with a 1 bar value of $T_{sol}^0 = 1373$ K and $dT_{sol}/dz = 3.3$ K km$^{-1}$, in rough agreement with experimental measurements (e.g., Hirschmann 2000).
In melting regions, the local melting rate, $q_m$, is calculated as $q_m(\vec x,t) = df/dt = (df/dz) u_z(\vec x,t)$, where $f$ is the degree of melting and $u_z=dz/dt$ is the upward convective flow velocity.
We use a constant value $df/dz=0.18 \%$ km$^{-1}$ for the adiabatic melt production per kilometer of upwelling \citep{PhippsMorgan2001}.
The local melting rate, $q_m$, is subsequently integrated over each melt column to give the surficial melt distribution.
We limit melt production to a maximum depth of 50 km, which results in a time-averaged melt production for a nominal Earth model (C150) consistent with Earth's present-day melt production of $\sim20$ km$^3$yr$^{-1}$ \citep{McKenzieBickle1988}.
Our choice of maximum depth affects the total melt production but has a small influence on the comparison between models or on surficial distributions of melt.
Because side boundaries promote vertical flow and unphysical concentration of melt at the substellar point, we calculate melting from models with a 360$^o$ opening angle.

\section{Tectonic Regimes}
\label{sct-res1}
\begin{figure} [H]
\begin{center}$
\begin{array}{c|c|c}
\includegraphics[width=5.2cm,angle=0]{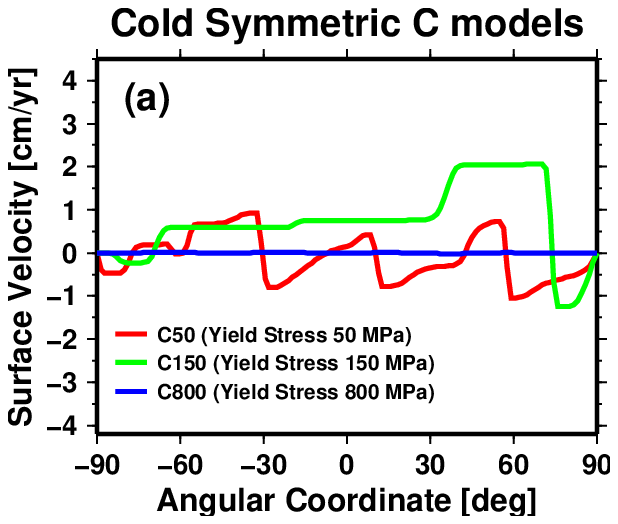}&
\includegraphics[width=4.9cm,angle=0]{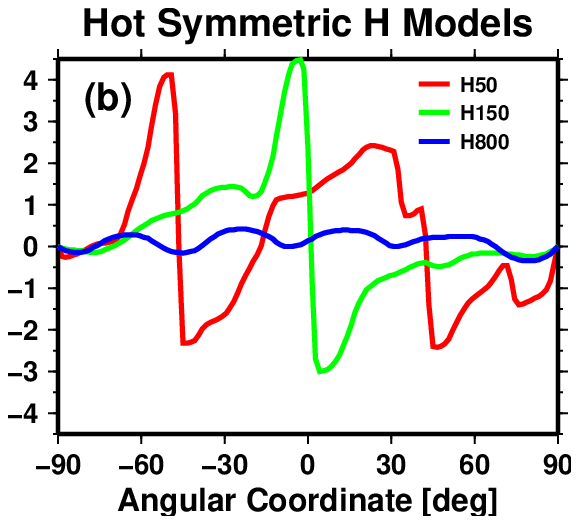}&
\includegraphics[width=4.9cm,angle=0]{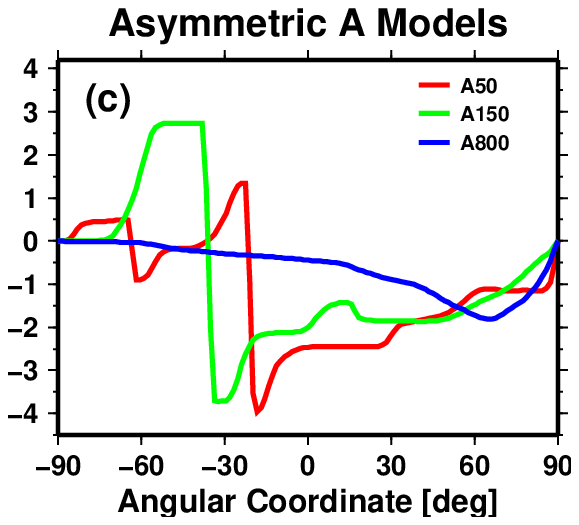}\\
\includegraphics[width=4.9cm,angle=0]{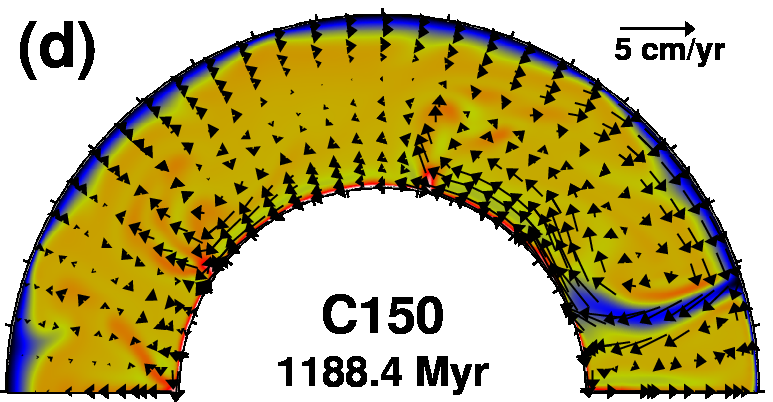}&
\includegraphics[width=4.9cm,angle=0]{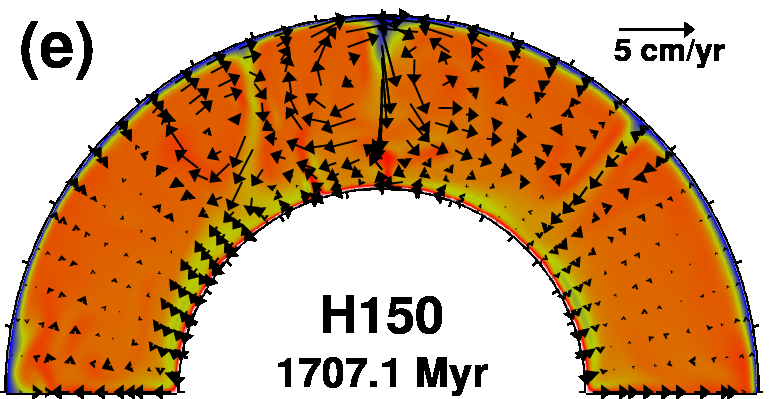}&
\includegraphics[width=4.9cm,angle=0]{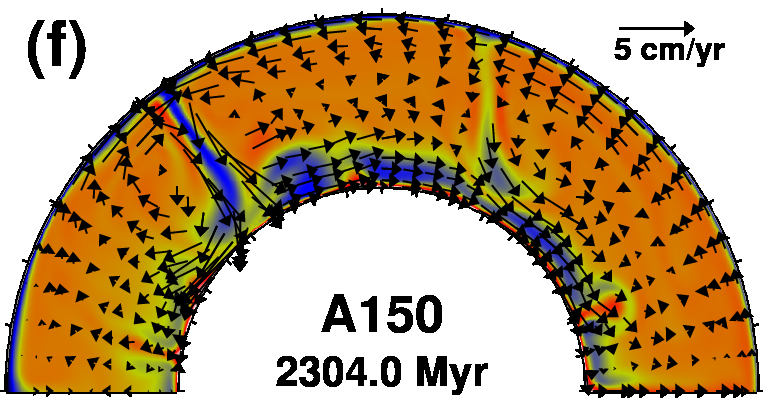}\\
\includegraphics[width=4.9cm,angle=0]{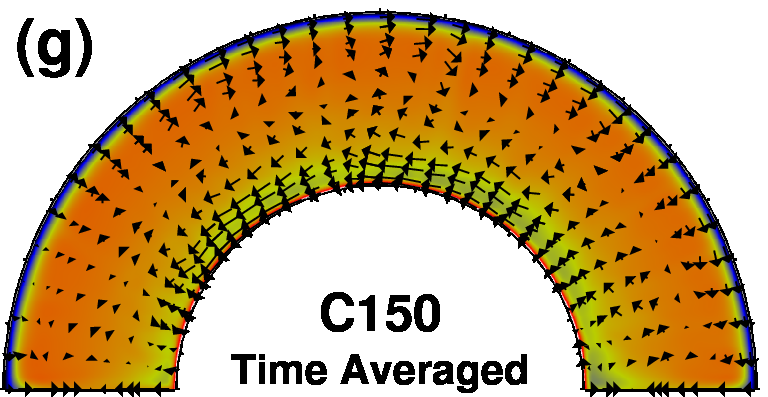}&
\includegraphics[width=4.9cm,angle=0]{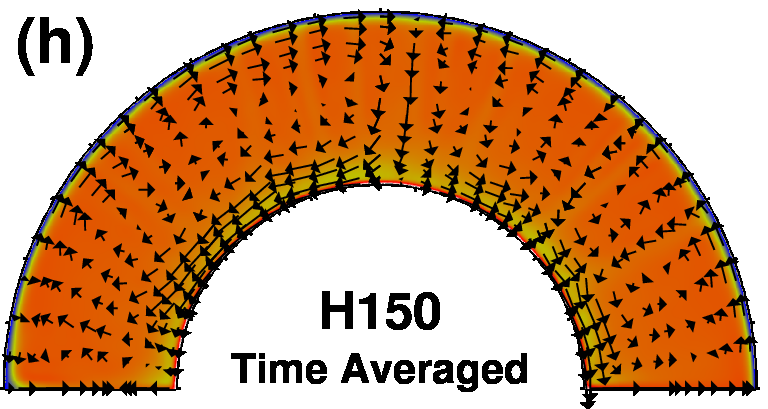}&
\includegraphics[width=4.9cm,angle=0]{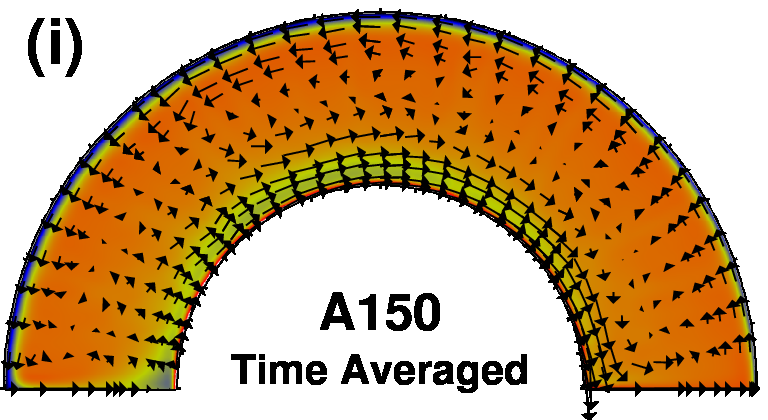}\\
&
\includegraphics[width=4.9cm,angle=0]{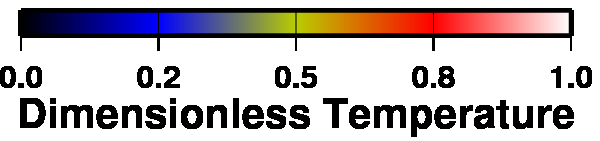}&
\\
\end{array}$
\caption{
Top row:
Surface velocities for
(a) cold symmetric C-models,
(b) hot symmetric H-models,
(c) hot asymmetric A-models
(see Section 2 for description).
For all model types, 3 cases are shown with different yield stresses
$\sigma_y$ of 50 MPa (red curve),
$150$ MPa (green curve), and
$800$ MPa (blue curve).
Middle row (d-f): Snapshots of dimensionless potential temperature and convective flow velocity for models with $\sigma_y$ = 150 MPa that correspond to the green curves in the top row.
Bottom row (g-i): Time-averaged values of the same quantities and models as in the middle row.
\label{fig-convect} }
\end{center}
\end{figure}

\begin{figure} [H]
\begin{center}$
\begin{array}{c}
\includegraphics[width=7.5cm,angle=0]{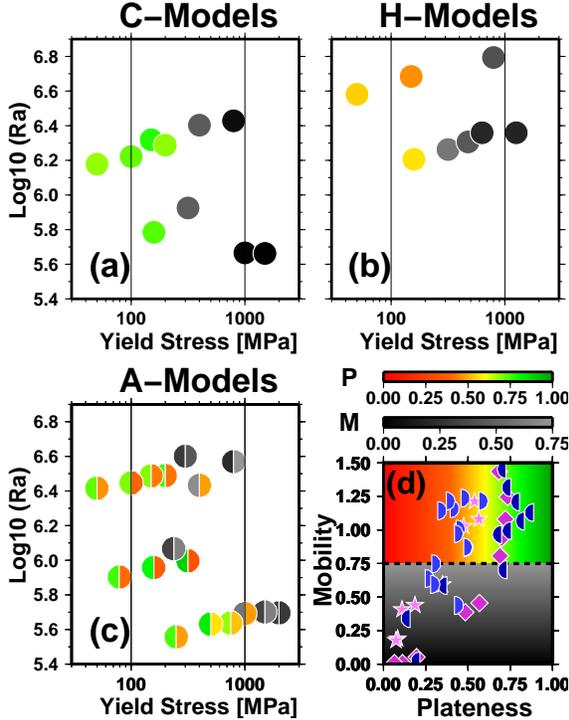}\\
\end{array}$
\caption{
Diagnostics of tectonic regime (plateness, $P$, and mobility, $M$) for models with different Rayleigh number and yield stress.
(a) "Cold symmetric" C-models,(b) "Hot symmetric" H-models, and
(c) "Hot asymmetric" A-models, for which both day (right) and night (left) sides are shown.
(d) All models in $P$-$M$ domain space:C-models (dark purple diamonds),
H-models (light purple stars),night side of A-models (dark blue left semi-circles), and
day side of A-models (light blue right semi-circles).Color coding in (a-c) is for $P$-values if $M$$\ge$0.75 and for $M$-values if $M$$<$0.75, as indicated by the color bars above frame (d).
\label{fig-domainplot} }
\end{center}\end{figure}

C-models show a variety of convection regimes for a progressive increase of the yield stress values $\sigma_y$.
At $\sigma_y \leq 50$ MPa, continuous yielding prevents the formation of stiff surface plates.
Instead, a mobile lid style of convection, characterized by diffuse surface deformation, occurs (Figure \ref{fig-convect}a, red curve).
At $\sigma_y \sim 150$ MPa, surface deformation is approximately plate-like with stiff surface plates separated by narrow regions of concentrated deformation.
Surface plates exhibit approximately piecewise constant velocities (Figure \ref{fig-convect}a, green curve) and converge toward subduction-like downwellings while diverging at localized spreading centers (Figure \ref{fig-convect}d).
A further increase of the yield stress produces a more time-dependent solution with alternating periods of surface mobility and stagnation.
For sufficiently high yield stress (800 MPa), surface mobility diminishes and stagnant lid convection persists (Figure \ref{fig-convect}a, blue curve).
To further demonstrate tectonic regimes, we show plateness $P$ and mobility $M$ in Figure \ref{fig-domainplot}.
For the C-models (Figure \ref{fig-domainplot}a), plateness increases to $P \sim 0.75$ with increasing yield stress at high mobility ($M\sim 1-1.5$) until the lid mobility rapidly decreases ($M\sim0$) around $\sigma_y \sim 250$ MPa (and plateness becomes irrelevant).

For the H-models, coherent lithospheric plates do not develop because low viscosities near the warm ($T_S$ = 759 K) surface promote viscous deformation and suppress pseudo-plastic yielding, consistent with the predictions of \citet{Lenardic_etal2008}.
Surface velocities are greater and have a more diffuse distribution than in the C-models (Figure \ref{fig-convect}b).
Due to increased lithospheric mobility (e.g. model H150, Figure \ref{fig-convect}e) plateness is consistently lower for the H-models ($P \leq 0.5$, Figure \ref{fig-domainplot}b) compared to the C-models ($P \sim 0.75$, \ref{fig-domainplot}a).
As $\sigma_y$ increases, stagnant lid prevails but plate-like behavior is not observed (Figure \ref{fig-domainplot}b).

\end{multicols}

\begin{multicols}{2}
For the A-models, a marked ($\Delta T_S$ = 800 K) temperature contrast causes global-scale asymmetries in tectonic regime.
Near the substellar point, high surface temperatures induce convective upwelling  and diffuse surface velocities (Figure \ref{fig-convect}c and \ref{fig-convect}f).
Away from the substellar point, plate-like behavior occurs and surface plates preferentially move toward the antistellar point (Figure \ref{fig-convect}f).
A global scale dichotomy is reflected in the tectonic regimes diagnosed separately for the day and night sides (Figure \ref{fig-domainplot}c).
The hot day side is consistently characterized by mobile lid convection with diffuse deformation, similar to the H-models, while the cold night side exhibits plate-like behavior for a large parameter space, similar to the C-models (Figure \ref{fig-domainplot}a).

\section{Patterns of Mantle Convection}
Flow patterns in the A-models are characterized by convective upwelling near the hot sub-stellar point, near-surface flow from the hot day side to the cold night
 side where most downwellings occur, and a deep mantle return flow toward the day side (Figure \ref{fig-convect}f).
An asymmetric degree 1 pattern of mantle flow persists, with convection cells that occupy the entire half-mantle domain (Figure \ref{fig-convect}i).
For the C- and H-models with uniform surface temperature, convective downwellings are more randomly distributed across the domain and this results in less persistent convective flow (Figure \ref{fig-convect}g and \ref{fig-convect}h).
As a diagnostic for persistent mantle flow, we use the RMS value of the time-averaged flow velocity normalized by the time average of the RMS flow velocities, or $\beta = (\overline{v})_{RMS}/\overline{v_{RMS}}$, where overlines indicate time-averaged values.
The persistent flow in model A150 is reflected by relatively high time-averaged convective flow velocities ($\beta$ = 0.53), compared to models with a uniform surface temperature, C150 ($\beta$ = 0.41) and H150 ($\beta$ = 0.43).

\begin{figure} [H]
\begin{center}$\begin{array}{c}\includegraphics[width=7.cm,angle=0]{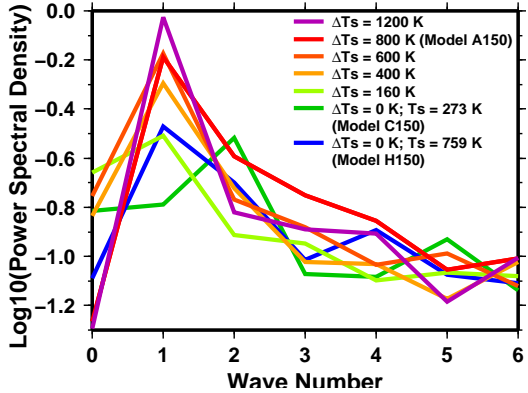}\\
\end{array}$
\caption{Power spectral density of lateral convective flow velocity.
Models with different surface temperature contrasts, $\Delta T_S$, and surface temperatures, $T_S$, are compared.
\label{fig-spectra} }
\end{center}
\end{figure}

To estimate the surface temperature contrast $\Delta T_S$ that is required for maintaining asymmetric mantle flow, we compare power spectra of time-integrated lateral convective flow velocities for models with different $\Delta T_S$ (Figure \ref{fig-spectra}).
For surface temperature contrasts $\gtrsim$400 K, a dominant degree 1 signal reflects persistent convection cells with upwelling at the substellar point and downwelling at the antistellar point.
A systematic degree 1 signal is not discernible for models with $\Delta T_S$$<$400 K and this reflects more randomly oriented convection cells with a richer variety of length scales.

\section{Rate and Distribution of Volcanism}
\label{sct-melt}
\begin{figure} [H]
\begin{center}$
\begin{array}{c}
\includegraphics[width=7.cm,angle=0]{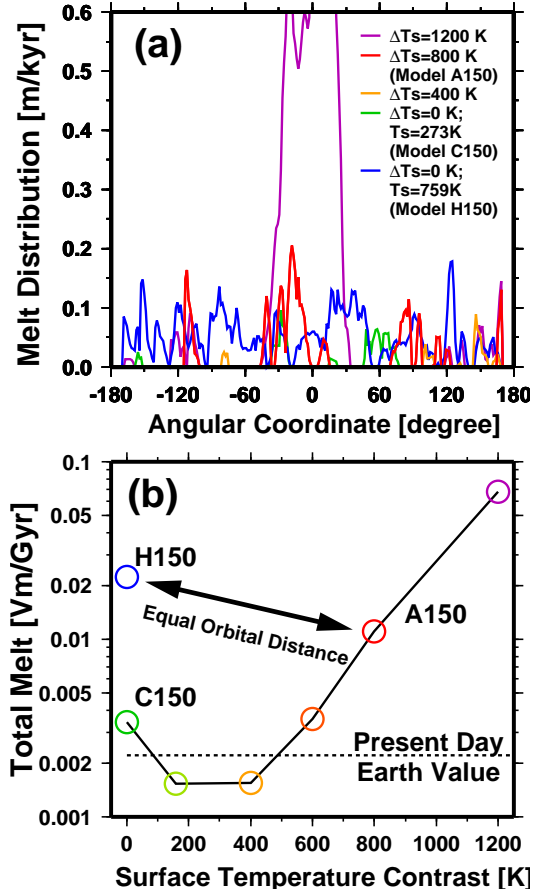}\\
\end{array}$
\caption{
(a) Distributions of time-averaged melt production.
Surface temperature contrasts, $\Delta T_s$ and surface temperatures, $T_S$, for different models are color-coded as shown in the legend.
The purple curve peaks at 1.18 m kyr$^{-1}$.
(b) Total melt production, shown as a processing rate in mantle volume per billion years.
The dashed line shows Earth's present-day melt production of $\sim$20 km$^3$ yr$^{-1}$, corresponding to $\sim$0.002 V$_m$ Gyr$^{-1}$.
\label{fig-melt} }
\end{center}
\end{figure}

For surface temperature contrasts $\Delta T_S\lesssim$400 K, melt rates show an uneven distribution without a coherent global pattern (Figure \ref{fig-melt}a, green and orange curves), and a total melt production similar to the present-day Earth value of $\sim$20 km$^{3}$ yr$^{-1}$ \citep{McKenzieBickle1988} (Figure \ref{fig-melt}b).
Above $\Delta T_S\sim$400 K, melt occurs preferentially within upwellings near the hot substellar point (Figure \ref{fig-melt}a, red and purple curves).
Due to more persistent day side melting, the total melt production rises above the present day Earth value by a factor of $\sim$5 for $\Delta T_S$ = 800 K and $\sim$30 for $\Delta T_S$ = 1200 K (Figure \ref{fig-melt}b, red and purple circles, respectively).

For planets at the same orbital distance, more than double the amount of melt is produced in the model with uniform surface temperature H150 (Figure \ref{fig-melt}b, blue circle) than in the asymmetric model A150 (Figure \ref{fig-melt}b, red circle).
Melting is more vigorous for model H150 because diffuse surface deformation occurs globally, and is less vigorous for model A150 because deformation occurs diffusely only near the substellar point.

\section{Discussion and Concluding Remarks}\label{sct-disc}
\begin{figure} [H]
\begin{center}$
\begin{array}{c}
\includegraphics[width=7.5cm,angle=0]{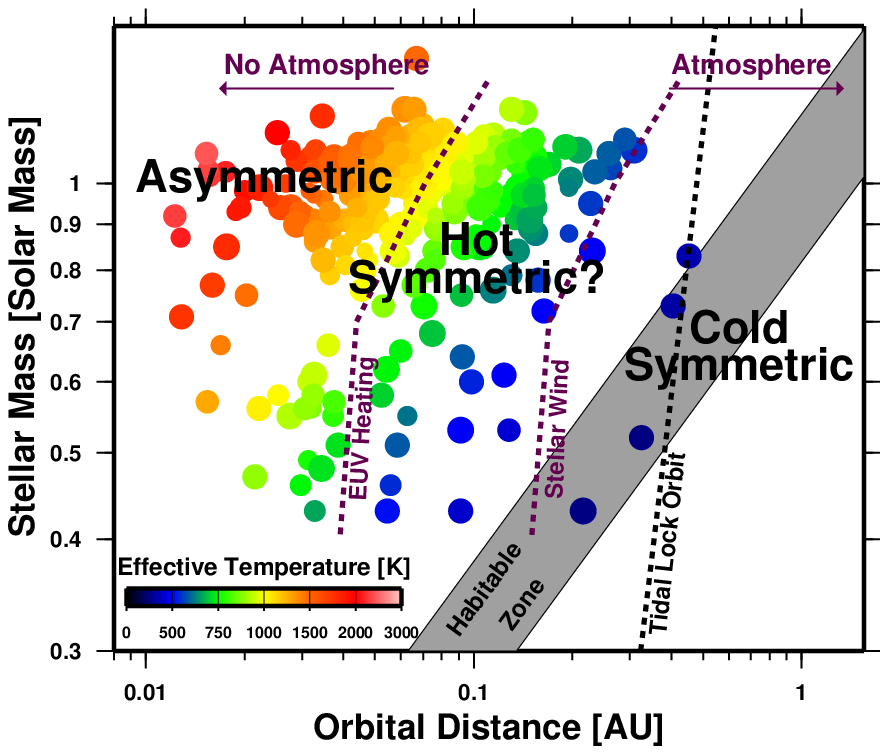}\\
\end{array}$
\caption{
Possible occupation of planets in orbital distance - stellar mass domain space.
Purple dotted curves show atmospheric escape estimates for EUV heating for a $6 M_E$ planet \citep{Tian2009} and stellar wind erosion for an Earth-sized planet \citep{Lammer_etal2008}.
Tidal locking is likely at distances smaller than the tidal lock orbit (black dotted line, \citep{Kasting_etal1993}).
Colored dots show effective temperatures, $T_{eff}$, for Kepler candidate exoplanets with $R<2R_E$ \citep{Borucki_etal2011}.
For planets with a substantial atmosphere, uniform surface temperatures are expected.
Planets without an atmosphere can have estimated substellar temperatures of $T_{substellar}$$\sim$$\sqrt{2}T_{eff}$ and a cold night side.
The theoretical habitable zone lies between the 273 and 373 K isotherms where liquid water can be maintained.
\label{fig-EPtemps} }
\end{center}
\end{figure}

Many recently-discovered exoplanets inhabit close-in orbits and this results in high (100's to 1000's K) effective temperatures (Figure \ref{fig-EPtemps}).
Planets closer than 0.5 AU are likely tidally locked (Figure \ref{fig-EPtemps}, black dotted line) \citep{Kasting_etal1993} and synchronously rotating around their parent stars.
At distances of $<$0.1 AU, Earth-mass planets are unlikely to retain an atmosphere, due to atmospheric loss either by extreme ultraviolet (EUV) heating \citep{Tian2009} or stellar wind erosion \citep{Lammer_etal2008} (Figure \ref{fig-EPtemps}, purple dashed curves).
These boundaries move outward for smaller planets that orbit around larger stars.
Absence of a substantial atmosphere results in high substellar temperatures, approximated as $T_{subst}=\sqrt{2} T_{eff}$, and the night side remains cold.
Thus, hot A-model behavior is expected for close-in orbits, and our results demonstrate that a persistent hemispheric surface temperature contrast $\gtrsim$400 K can maintain a degree 1 pattern of mantle convection in which the planetary surface moves preferentially toward subduction zones on the cold night side.
These planets should exhibit an inhospitable day side with vigorous volcanism and a cold night side that allows for more Earth-like tectonics and plate-like behavior.
Melt production increases with increasing surface temperature contrast, and at 0.13 AU the calculated total melting rate is $\sim$5 times higher than for a cool planet at 1 AU where plate tectonic behavior occurs globally (C-models).

At intermediate distances ($\sim$0.1 AU), an atmosphere is more likely to persist and atmospheric redistribution of heat is expected to produce a uniform hot surface.
For such planets, our H-model results demonstrate that mobility of the lithosphere prevents the formation of coherent plates and the emergence of Earth-like plate tectonics.
Compared to an asymmetric model at the same orbital distance, melt production is enhanced by a factor $>$2, and occurs globally across a surface that is characterized by diffuse deformation.
Our results suggest the possbility of different feedback mechanisms.Volcanic outgassing is responsible for the formation of secondary planetary atmospheres.
Therefore, asymmetric conditions can only be sustained if the atmosphere is continuously eroded.
If, instead, an atmosphere is retained, heat redistribution promotes mobile lid convection with diffuse deformation and further increases global volcanic outgassing.
This suggests a positive feedback in favor of a thick atmosphere, unless another mechanism modifies the atmospheric balance.
For example, no melt production is calculated for one-plate model planets, a possibility for potential close-in equivalents of present-day Venus.
Atmospheres would be unprotected against solar wind erosion if a magnetodynamo is deactivated due to a transformation from plate tectonics to a regime that is less efficient at cooling the mantle and core \citep{Buffett2002,ChristensenTilgner2004,Gaidos_etal2010}.
Substantial atmospheric loss would allow for a negative feedback and such planets could fluctuate between symmetric and asymmetric end- member scenarios or reach an intermediate equilibrium, depending on the timescale of atmospheric loss relative to mantle thermal evolution.
The dependence of tectonic regimes on $\sigma_y$ for our nominal Earth models is  consistent with previous studies performed using 2-D cartesian, 3-D cartesian, and 3-D spherical geometries \citep{MoresiSolomatov1998,Tackley2000b,Richards_etal2001,VanHeckTackley2008,FoleyBecker2009}.
As in the above-mentioned studies plate-like behavior occurs at yield stresses lower than classical estimates for dry oceanic lithospheric strength by up to $\sim$1000 MPa \citep{Kohlstedt_etal1995}, but weak fault zone fabric may account for this discrepancy \citep{MooreRymer2007,Escartin_etal2008}.
The weak dependence of tectonic regimes on $Ra$ shown in Figure \ref{fig-domainplot}a is in agreement with numerical results of \citet{FoleyBecker2009} and scaling laws of plate tectonic convection by \citet{Korenaga2010a,Korenaga2010b}.
Because a higher $Ra$ is expected for more massive planets, this low sensitivity to $Ra$ suggests that our results can be applied to planets of various sizes.

Convection in massive "super-Earth" mantles may be influenced by their more extreme pressure and temperature conditions.
For example, mineral physics calculations suggest a viscosity decrease of 2 to 3 orders of magnitude for the deep mantles of super-Earths \citep{Karato2011}, in favor of vigorous convective overturn.
High mantle pressure allows for mineral phase transformations that do not occur in Earth's mantle \citep{Umemoto_etal2006} but which can strongly influence the dynamics of super-Earth mantles \citep{VandenBerg_etal2010}.

Although important questions remain, our simulations demonstrate the strong influence that surface temperature contrasts exert on mantle convection, surface tectonics, and volcanism for close-in rocky exoplanets.
Distinct scenarios are likely associated with variations in albedo, volcanism, and atmospheric content that may become astronomically detectable in the future.
\end{multicols}

\acknowledgments{
This research was supported by the NSF grant EAR-0914712 and NASA grant NNX10AI90G.
We thank Michael Manga, Edwin Kite, Ray Pierrehumbert, and Maxim Ballmer for helpful discussions and an anonymous reviewer for constructive comments.
}


\end{document}